\documentclass{ws-rv9x6}
\usepackage{subfigure}   
\usepackage{ws-rv-thm}   
\usepackage{ws-rv-van}   
\makeindex

\begin{document}


\begin{center}
\bf \large {Nucleon-anti-nucleon intruder state of
\\ Dirac equation for nucleon in deep scalar potential well}
\end{center}

\author{T.T.S. Kuo$^{a}$\footnote{thomas.kuo@stonybrook.edu},
T.K. Kuo $^b$ , E. Osnes $^{c}$,  S. Shu $^{a,d}$}
\vskip 0.3cm
\address{ $^a$Department of  Physics and Astronomy,
 Stony Brook University,\\
Stony Brook, NY 11794, USA}
\address{ $^b$ Physics Department,  Purdue University,
W. Lafayette, IN  47907, USA\\}
\address{ $^c$ Institute of  Physics, University of Oslo,
NO 0316 Oslo, Norway}
\address{ $^d$Department of Physics and Electronic Science, Hubei Univ.,
\\ Wuhan 430062, China}


\begin{abstract}
We solve the Dirac radial equation for a nucleon in a
scalar Woods-Saxon  potential well of depth $V_0$ and radius
$r_0$. A sequence of values for the depth and radius are considered.
For shallow potentials with $ -1000 MeV\lesssim V_0 < 0$
the wave functions for the positive-energy states $\Psi _+(r)$ are
dominated by their nucleon component $g(r)$.
But for deeper potentials with $V_0 \lesssim -1500 MeV $
the $\Psi_+(r)$s  begin to have dominant
anti-nucleon component
$f(r)$. In particular,  a special intruder state enters
with wave function $\Psi_{1/2}(r)$ and energy $E_{1/2}$.
We have considered several  $r_0$ values between 2 and 8 fm.
For $V_0 \lesssim -2000 MeV$
 and  the above $r_0$ values, $\Psi _{1/2}$ is the only bound
positive-energy state and  has its  $g(r)$ closely equal to $-f(r)$,
both having a narrow wave-packet shape centered around $r_0$.
The  $E_{1/2}$ of this state is practically independent
of $V_0$ for the above $V_0$ range  and obeys
 closely the relation
$E_{1/2}=\frac{\hbar c}{r_0}$.

\end{abstract}


\section{Introduction}\label{introd}
   Intruder states have played an important role in
nuclear structure physics. Two of us (TTSK and EO) first
learned of this subject from Gerry Brown when we were his postdocs
  at respectively Princeton and  Nordita some
long time ago. Gerry has taught, inspired and helped both of us
a great deal; we are deeply grateful to him and remember well
the  pleasant time we had with him.

The nuclear shell model is a most succesful model for
nuclear structure. A desirable feature  is its
large energy gap between major shells. For example, the large
energy gap between the $1s0d$ and $0p$ shells allows us to treat
 the nucleus $^{18}O$  using a small $2p0h$ model
space, namely treating it simply  as two $0s1d$ valence
nucleons  outside a closed $^{16}O$ core. Pioneered  by Gerry,
a large number of shell-model studies of nuclei using realistic
nucleon-nucleon interactions have been carried out.
\cite{kuobrown,brownkuo67,brownbook71,jensen,coraggio09,
brown2010,dong14,kuo16}
 The importance of  intruder states
in  model-space effective interactions  has been
investigated by Schucan and Weidenmueller and others
. \cite{weiden73,ellis77}

  In Fig.1 we display the results of a typical calculation
of $^{18}O$ \cite{dong14}. In this calculation, a low-momentum
interaction $V_{lowk}$ is employed, and the shell-model effective
interaction has been calculated using three effective interaction
methods  denoted by LS, KK and EKKO \cite{dong14} in the figure.
As seen, calculations give only two
$2^+$ states while experiments have three. The missing one
is an 'intruder' state whose wave function is mainly outside
 the $2p0h$ space which is the model space used for
such calculation. The energies of the unperturbed
$4p2h$ states are much higher than the $2p0h$ ones. But when the
interaction is strong, some $4p2h$ states are pushed down,
forming an intruder $2^+$ state whose energy is as low
as those states which  are dominated by the $2p0h$ components.
The presence of the intuder state is an indication that  the
Fermi sea ($^{16}O$-core) is no longer closed when the
interaction is strong.

In the present short contribution dedicated to the memory
of Gerry, we shall study
 the Dirac equation for a nucleon  in a strong scalar potential
well and discuss its possible intruder states. For a free nucleon,
the negative-energy Dirac sea is closed. But when it is subject
to a strong external field, it may   be no longer closed
and in anlogy to the nuclear case described above,
give rise to nucleon-anti-nucleon intruder states.

\begin{figure}[h]
\scalebox{0.45}{\includegraphics[angle=0]{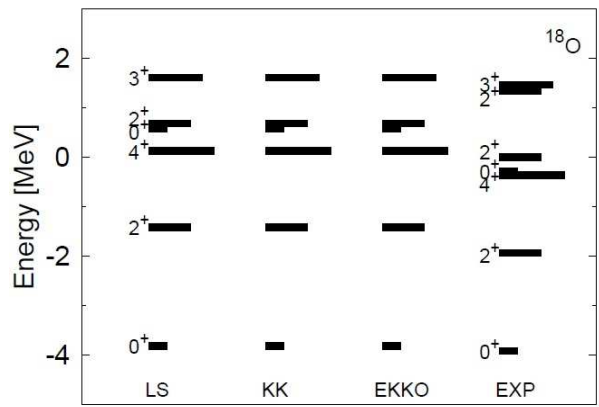}}
\caption{Intruder states in low energy spectrum of $^{18}O$}
\end{figure}

   Before discussing
our present work, we would like to
mention briefly Gerry's style of  physics; he was fond
of and successful in using simple models to
bring out the  simple physics
of complicated problems. He often said 'This is too complicated;
I can not check it with my 5-dollar  calculator'.
The Brown-Bolsterli schematic model for
nuclear giant dipole resonance is a well known  example
\cite{brown59}. The second-order core polarization
diagram for nuclear effective interaction is another
 example, and has played  an essential role
in microscopic nuclear structure studies
\cite{kuobrown,brownkuo67,brownbook71,jensen,coraggio09,
brown2010,dong14,kuo16}.
How would the properties of hadrons change with their surrounding
medium is a quite complicated problem. But the Brown-Rho
scaling \cite{brownrho91,brownrho04} provides a remarkably
simple answer, namely  $m^*/m~=~1-Cn/n_0$ where $m^*$ and $m$
are respectively the in-medium and in-vacuum hadron mass.
The medium density is $n$ and the nuclear matter saturation
density $n_0$. For low densities, $C \simeq$0.2. Extensive
applications of the
Brown-Rho scaling on nuclear matter, neutron stars
and the long life-time $\beta$-decay of $^{14}C$
 have been carried out
\cite{holtbrs07,holtc1408,siu09,dong09,dong11,dongnewbr13,
paenghls16}.

  In the following we shall first describe some details for
solving the Dirac radial equation with a scalar central potential
 $V_s$  (section 2). To mimic the extremly strong potential
which may be present in highly compacted stellar objects, we shall
consider the Dirac equation for a nucleon in super-deep scalar potentials
in section 3. The resulting
nucleon-anti-nucleon intruder states will be discussed.
In the final section 4 we present a summary and conclusion.

  In order to elucidate our calculation by a simple pedagogical
example, we have repeated the calculation using a one-dimensional
Woods-Saxon potential; this is included in an Appendix.

\section{Dirac radial equation with scalar potential well}\label{dirac}
 We consider the Dirac equation for  a nucleon
\begin{equation}
[c({\bf\alpha \cdot p}) +{\bf\beta} (mc^2+V_s) +V_v]\psi=E \psi
\end{equation}
where $m$ denotes its mass, and $V_s$ and $V_v$ respectively
the scalar and vector potential.
 Its radial equation \cite{nogami90,landau96,silbar11} is
\begin{eqnarray}
\frac{dg(r)}{dr}&=& -\frac{k}{r}g(r)
+\frac{1}{\hbar c}
   [E-V_v(r)+V_s(r)+mc^2]f(r)  \nonumber \\
\frac{df(r)}{dr}&=& +\frac{k}{r}f(r)
  -\frac{1}{\hbar c}[E-V_v(r)-V_s(r)-mc^2]g(r)
\end{eqnarray}
with  
\begin{eqnarray} k=\left \{
\begin{array}{lc}
-(l+1),\ \ & \ \ ~j=l+1/2  \\
  l, \  \ & \  \ j=l-1/2
\end{array}
\right.
\end{eqnarray}
and
\begin{equation}
\psi _{jlm}=\frac{1}{r}\left[ \begin{array}{c}
     ig(r)y^j_{lm}\\  -f(r)y^j_{l'm}
                   \end{array} \right]
\end{equation}
where $l'=2j-l$.
 In the above, we use  $mc^2$=938 MeV and $\hbar c$=197.3 MeV-fm.
We shall consider in the present work the $l=0$ and $j=1/2$
state (namely k=-1).

To determine the bound-state eigenvalues of the above Dirac radial
equations, a standard procedure is to integrate  from r=0
to $r_{end}$ in two portions: (I) from $r=0$ to $r_{match}$
obtaining wave functions
$g_I$ and $f_I$, and (II) from $r_{end}$ to $r_{match}$ obtaining
$g_{II}$ and $f_{II}$, where $0< r_{match} <r_{end}$ .
For an energy variable $\omega$
equal to a bound-state  eigenvalue
the logarithmic boundary conditions at $r_{match}$ are satisfied, namely
\begin{equation}
 \frac{d}{dr} log(g_I(\omega,r_{match}))
 =\frac{d}{dr} log(g_{II}(\omega,r_{match})),~\omega=E_1,E_2,\cdots,
\end{equation}
and
\begin{equation}
 \frac{d}{dr} log(f_I(\omega,r_{match}))
 =\frac{d}{dr} log(f_{II}(\omega,r_{match})),~\omega=E_1,E_2,\cdots.
\end{equation}

Note that the above two conditions are theoretically equivalent,
but 'numerically' they are often not.
 High accuracy is generally needed for calculating the above
quantities, and when this is not met they
may give different and spurious eigenvalues.

 In the present work we employ the following alternative
matching condition  for determining  the
eigenvalues, namely
\begin{equation}
 \frac{f_I(\omega,r_{match})}{g_I(\omega,r_{match})}
 =\frac{f_{II}(\omega,r_{match})}{g_{II}(\omega,r_{match})}.
\end{equation}
 This single condition is 'theoretically'
equivalent to the previous two conditions, Eqs.(5) and (6).
But 'numerically' we have found it being considerably
more accurate and efficient.
 In our calculations, we shall mainly use this
 matching condition.  We shall also use both Eqs.(5) and (6)
and a visual inspection of the resulting wave functions
 to  double check our results.

 We consider a nucleon in a scalar
Woods-Saxon potential well
\begin{equation}
V_s(r)=\frac{V_0}{1+ e^{(r-r_0)/\delta}}
\end{equation}
with well depth  $V_0$.
We shall use a range of $V_0$ values from about $-50$ to minus
few-thousand MeV. (Possible connection of our choices with
dense stellar objects like black holes will be discussed in section 4.)
We shall also consider a range of $r_0$ values.
For convenience we shall use $\delta$=0.1 fm for all calculations.
The values
 for $r_{match}$ and $r_{end}$ are dependent on $r_0$.
For example for $r_0=$ 4 fm we use  $r_{end}$= 6 fm
 and $r_{match}$= 4.5 fm. For this case we have also used
$r_{match}$= 5.0 fm, with  results
 in very good agreement (to 4th decimal place) with the former.

  To illustrate our calcualtions, we first present some of our
results with a $V_0=-50$ MeV potential, $r_0$=4 fm and $\delta$
as given above.
As seen in Fig.2, the wave function of $E_{1+}$ is largely
 dominated by its nucleon component $g(r)/r$. But the wave
function for the negative energy $E_{1-}$ is dominated
by its anti-nucleon component $f(r)/r$ as seen in Fig. 3.
 In Fig. 4  we consider  the wave functions corresponding
of the $E_{2+}$ state of the same potential. They are still
dominated by $g(r)/r$ but have a slightly
larger $f(r)/r$ component
than $E_{1+}$. Note both of the wave functions in Fig. 4 have
a node.
 Our scheme for
ordering the positive energy states is
$E_{1+}<E_{2+}<E_{3+}< \cdots$
 and $\cdots <E_{3-}<E_{2-}<E_{1-}$ for the negative energy ones.

 These wave functions
and those to be displayed later are normalized without
the angular integration
factor $4\pi$, namely
\begin{equation}
\int (f(r)^2+g(r)^2)dr=1.
\end{equation}

  The above optical potential  is similar to the empirical
optical potentials used for a nucleon in a nucleus of mass number
$A\simeq$40 \cite{hodgson94,nogami90}. Our results indicate that
for such ordinary nuclear systems
the positive energy states of their nucleons are dominated
by their nucleon component $g(r)$.

\begin{figure}[h]
\scalebox{0.36}{\includegraphics[angle=-90]{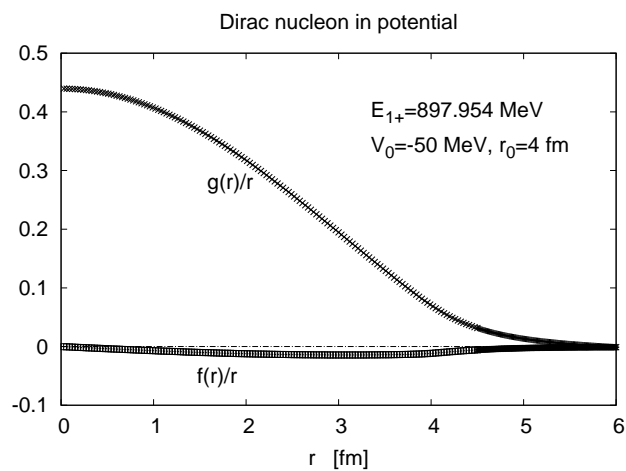}}
\caption{Dirac wave functions for $E_{1+}$=897.952 MeV and
$V_0$=-50 MeV.}\label{figure1}
\end{figure}

\begin{figure}[h]
\scalebox{0.36}{\includegraphics[angle=-90]{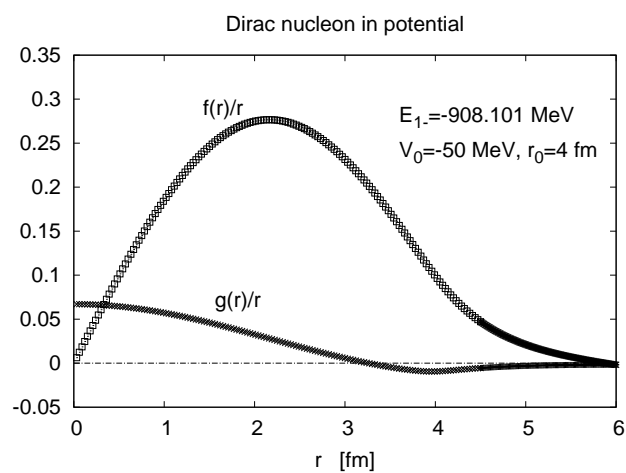}}
\caption{Same as Fig. 2 but for $E_{1-}$=-908.100 MeV.}\label{figure3}
\end{figure}

\begin{figure}[h]
\scalebox{0.36}{\includegraphics[angle=-90]{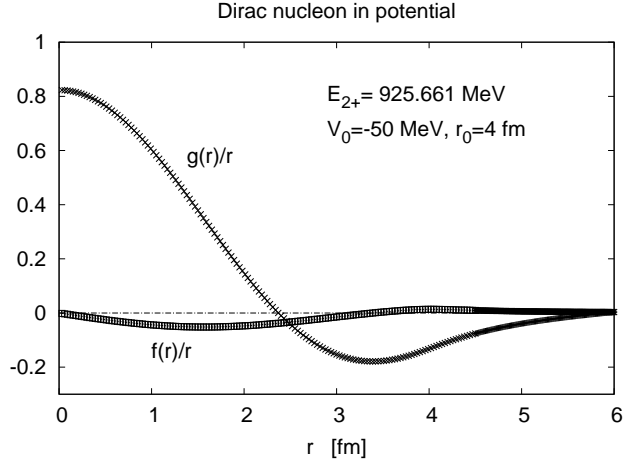}}
\caption{Same as Fig. 2 but for $E_{2+}$=925.661 MeV.}
\end{figure}

\section{Nucleon anti-nucleon level crossing and
intruder state}


\begin{table}[ht]
\tbl{Energies and anti-nucleon fraction
 $\langle f|f \rangle$ of
a nucleon in scalar Woods-Saxon potentials of depth $V_0$.
The widths of these potentials are all $r_0$=4 fm.
The subscripts $1+$ and $1-$ refer to respectively the lowest positive-
and highest negative-energy state. $V_0$ and $E$ are both in units
of MeV.}
{\begin{tabular}{@{}ccccc@{}} \toprule
  ~~$ V_0$ ~ &~~  $E_{1-}$ ~ & ~~ $\langle f|f \rangle_{1-}~ $ &
  ~~  $E_{1+}$~ &~~  $\langle f|f \rangle _{1+}$~\\ \colrule
  -50  &  -908.101 &  0.9908 &  897.954  & 0.004645\\
  -100 &  -861.325 &  0.9883  & 849.499 &  0.005851\\
  -300 &  -671.036 &  0.9780  & 654.283 &  0.01098\\
  -500 &  -485.047 &  0.9570  & 461.309 &  0.02180\\
  -700 &  -313.559 &  0.8594  & 275.994 &  0.05498\\
  -900 &  -194.461 &  0.6487  & 120.856 &  0.2011\\
 -1000 &  -191.321 &  0.3894  & 76.8122 &  0.3537\\
 -1100 &  -236.603 &  0.1854  & 59.8212 &  0.4574\\
 -1300 &  -397.210 &  0.05180 &  52.5618 &  0.4917\\
 -1500 &  -583.955  &  0.02299 &  50.7829  & 0.4960\\
 -1700 &  -777.233 &  0.01515 &  49.9197 & 0.4973\\ \botrule
\end{tabular}
}
\label{ra_tbl2}
\end{table}


\begin{figure}[h]
\scalebox{0.38}{\includegraphics[angle=-90]{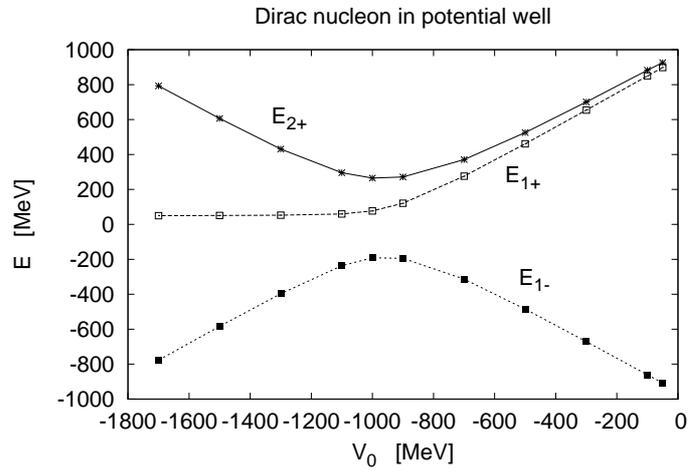}}
\caption{Energy levels of  nucleon in  potential wells
of different well depths $V_0$.}\label{figure3}
\end{figure}

\begin{figure}[h]
\scalebox{0.38}{\includegraphics[angle=-90]{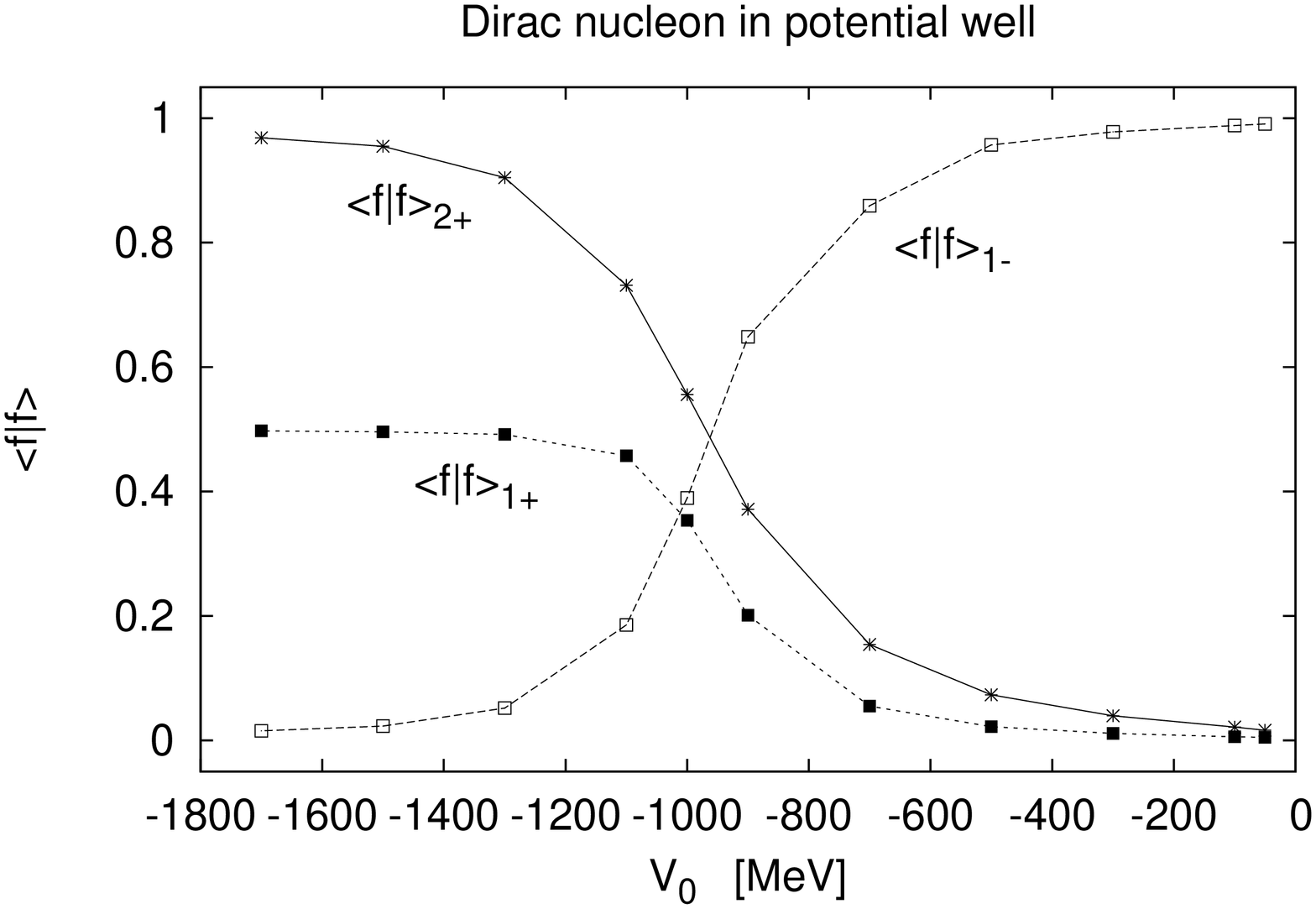}}
\caption{Anti-nucleon fractions of  nucleon in  potential wells
of different well depths $V_0$.}\label{figure3}
\end{figure}

 In  Table 1, we present our results for a Dirac nucleon
in the above potential well with different well depths $V_0$.
Before discussing our results, let us mention that the matching
conditions addressed earlier depends very sensitively
on the energy
variable $\omega$.
Thus, a very small change  in $\omega$ can
drastically change the wave functions
$f$ and $g$. As shown in Table 1,
we have determined the energies  $E$ very accurately
so as to satisfy the matching conditions. It is our hope that
 interested readers may check our numerical results.

Let us now discuss
the results  listed in Table 1. In Fig. 5 we plot the energies of
of the $1-$, $1+$ and $2+$ states versus $V_0$. As seen there is clearly
a level-crossing behavior; the nucleon level $E_{1+}$ descends
with $V_0$ whereas the anti-nucleon $E_{1-}$
level ascends. They 'cross'
at a crossing potential of $V_0\simeq -1000$MeV$\simeq -mc^2$.
(In fact they can not actually
cross for real $V_0$, but can do so for complex $V_0$.\cite{weiden73})
In Fig.6, we plot the corresponding anti-nucleon fractions
$ \langle f|f \rangle$. As the well becomes deeper,
 the anti-nucleon fraction of the $1-$ state drops
monotonically from $\sim 1$ at shallow well to $\sim 0$
at deep well of  $V_0 \simeq -2000 MeV$. In other words,
this anti-nucleon is 'transmuted' into a nucleon in the process.
We note that the crossing potentials for $E$ and
$\langle f|f \rangle$ are approximately the same.

\begin{figure}[h]
\scalebox{0.4}{\includegraphics[angle=-90]{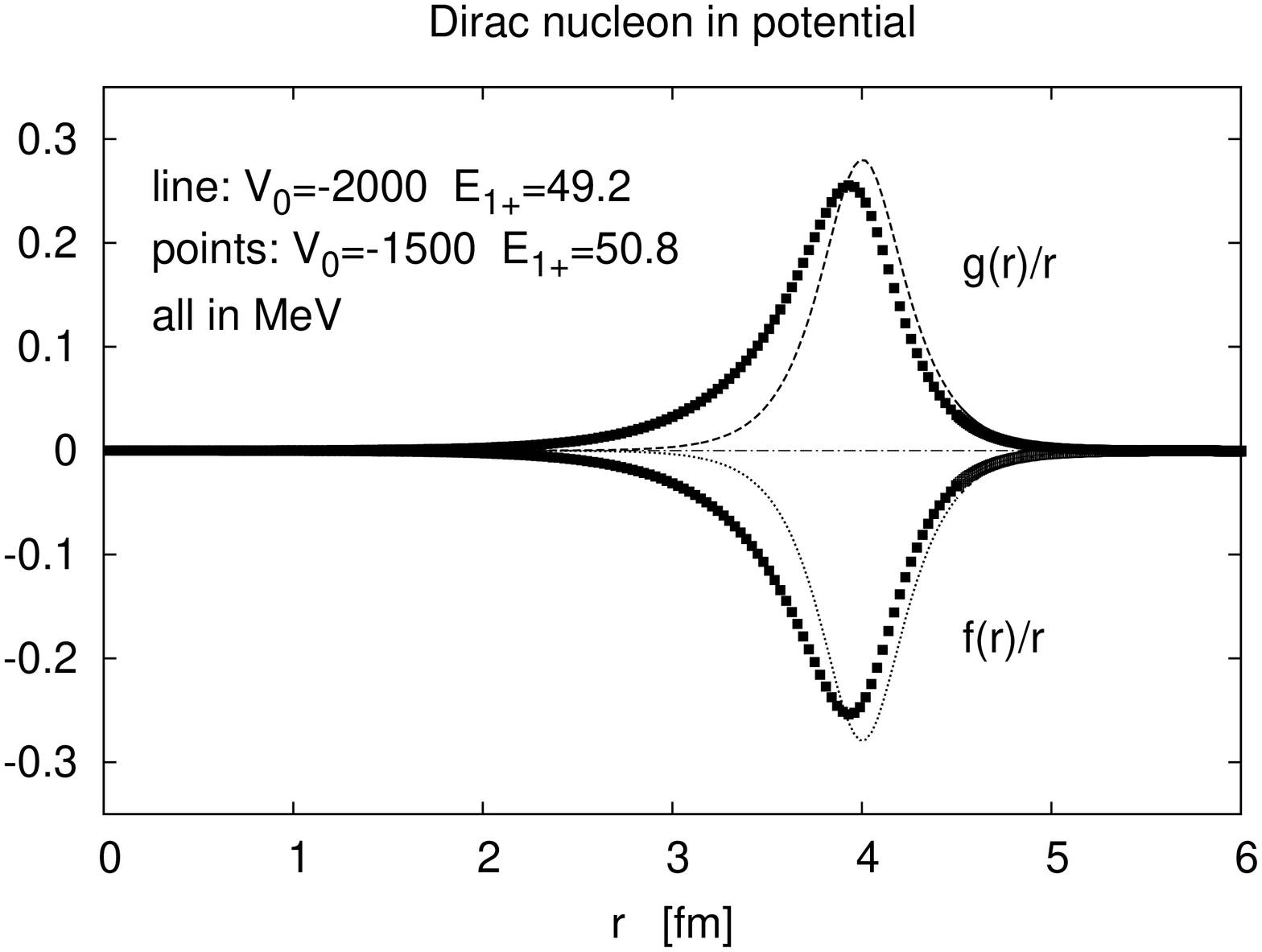}}
\caption{Wave functions of a  Dirac nucleon in a
super-deep potential well of depth $V_0=-1500$ MeV with energy
$E_{1+}$=50.7829 MeV. Those of the corresponding state
for $V_0$= -2000 MeV are also shown.}
\end{figure}

\begin{figure}[h]
\scalebox{0.4}{\includegraphics[angle=-90]{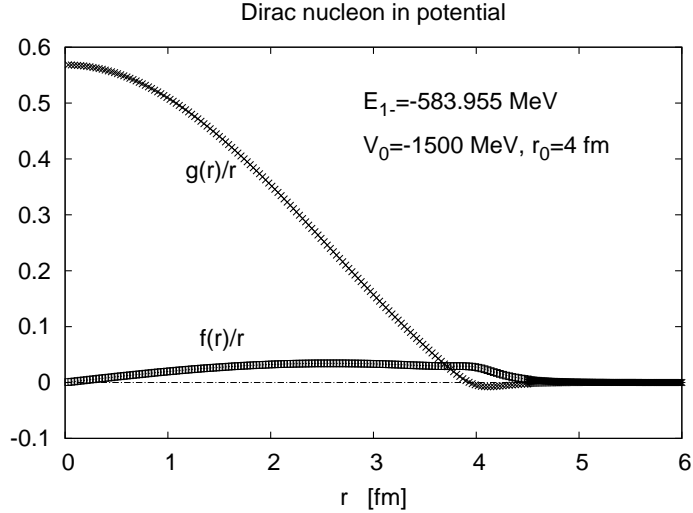}}
\caption{Same as Fig.7 but for energy $E_{1-}$=-583.955 MeV.}
\end{figure}

\begin{figure}[h]
\scalebox{0.43}{\includegraphics[angle=0]{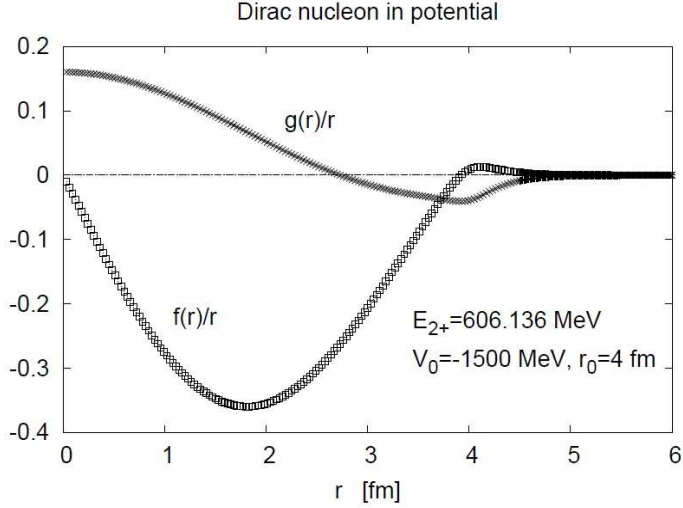}}
\caption{Same as Fig.7 but for energy $E_{2+}$=606.136 MeV.}
\end{figure}

  We now consider the evolution of the energy and anti-nucleon
fraction of the $1+$ state  in Figs. 5 and 6. As the well
becomes deeper, its energy drops till about $V_0 \simeq -1000$ MeV.
And until this strength is reached, this anti-nucleon fraction  keeps on
rising. But afterwards a plateau
of energy $\simeq 50$ MeV and $\langle f |f \rangle \simeq 0.50$
is reached.
It is of interest that this nucleon
 is approaching some sort of
'hybrid nucleon' composed of half nucleon and half anti-nucleon.
It may be noted that, as indicated by Fig. 5, this hybrid state
is the only bound state for $V_0 < \sim -2000$ MeV,
 while all other states
 are diverging outward having  energies $|E|>mc^2$.
 (By bound state
we mean a state of energy being $|E|< mc^2$.)

\begin{figure}[h]
\scalebox{0.4}{\includegraphics[angle=-90]{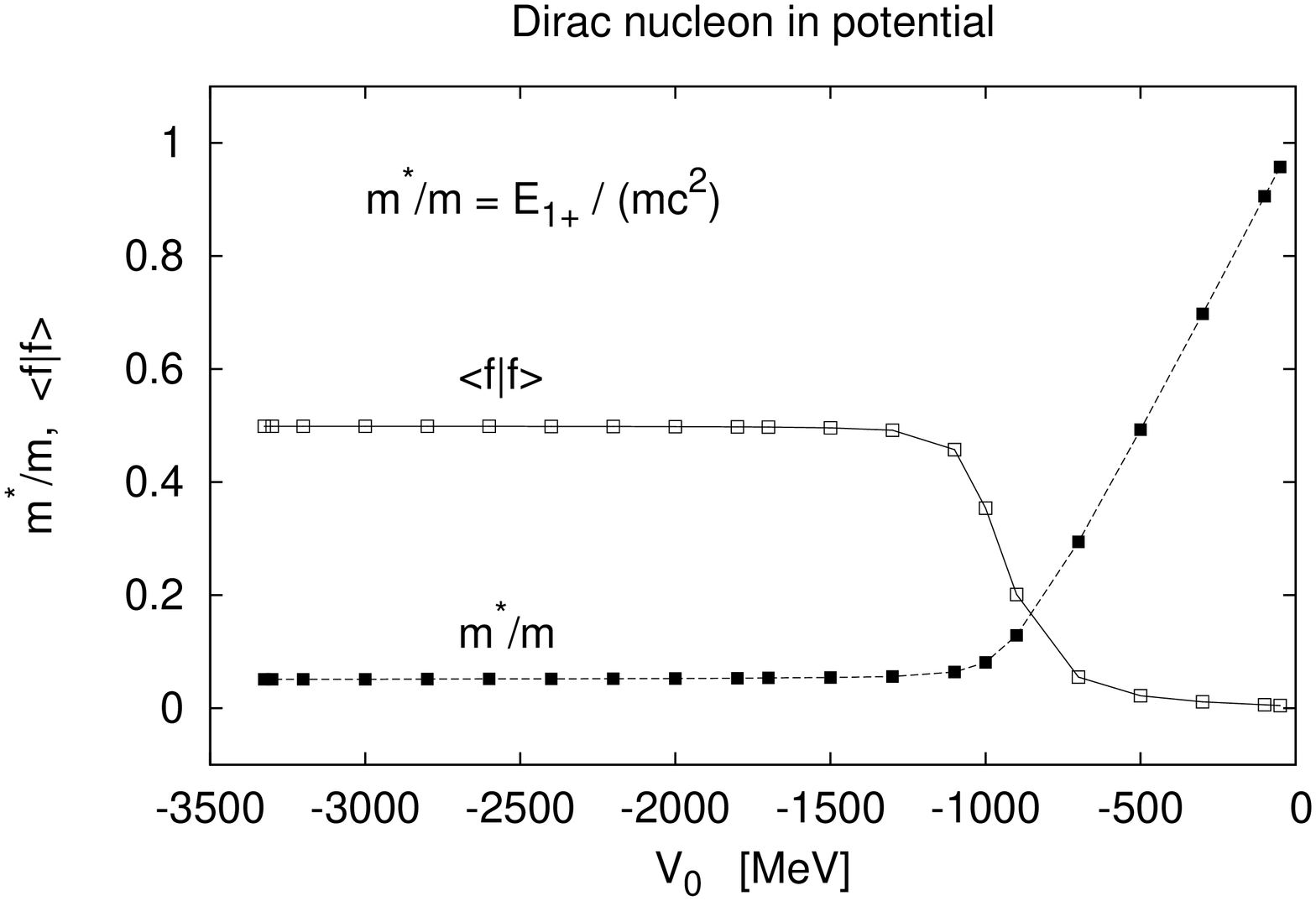}}
\caption{Evolution of the $E_{1+}$ energies and their anti-nucleon fractions
 with  potential
 depth $V_0$. A potential radius of $r_0$=4 fm is used
for  all calculations shown.}\label{figure3}
\end{figure}

  In Fig. 7 we display the wave function of such a hybrid nucleon
belonging to the $V_0=-1500$ MeV potential.
It is amusing that the radial distributions of its nucleon
and anti-nucleon components are almost identical,
 except for a sign change in the wave function.
Furthermore, this hybrid nucleon has the mass largely concentrated
near its surface, making it  a hollow nucleon. In this figure,
the corresponding wave function for $V_0=-2000$ MeV is also shown.
It has a similar Gaussian wave packet shape, but of narrower width.


 The wave function of the $1-$ anti-nuleon belonging to the same
-1500 MeV potential is displayed in Fig. 8. As seen, it has almost
no anti-nucleon content; it is a 'nucleon' evolved from
its parent anti-nucleon. The wave function of the $2+$ state of this
potential is shown in Fig. 9. Its $f(r)$ component
is much larger than the $g(r)$ one, indicating this positive energy
nucleon is actualy predominantly composed of anti-nucleon.


\begin{figure}[h]
\scalebox{0.4}{\includegraphics[angle=-90]{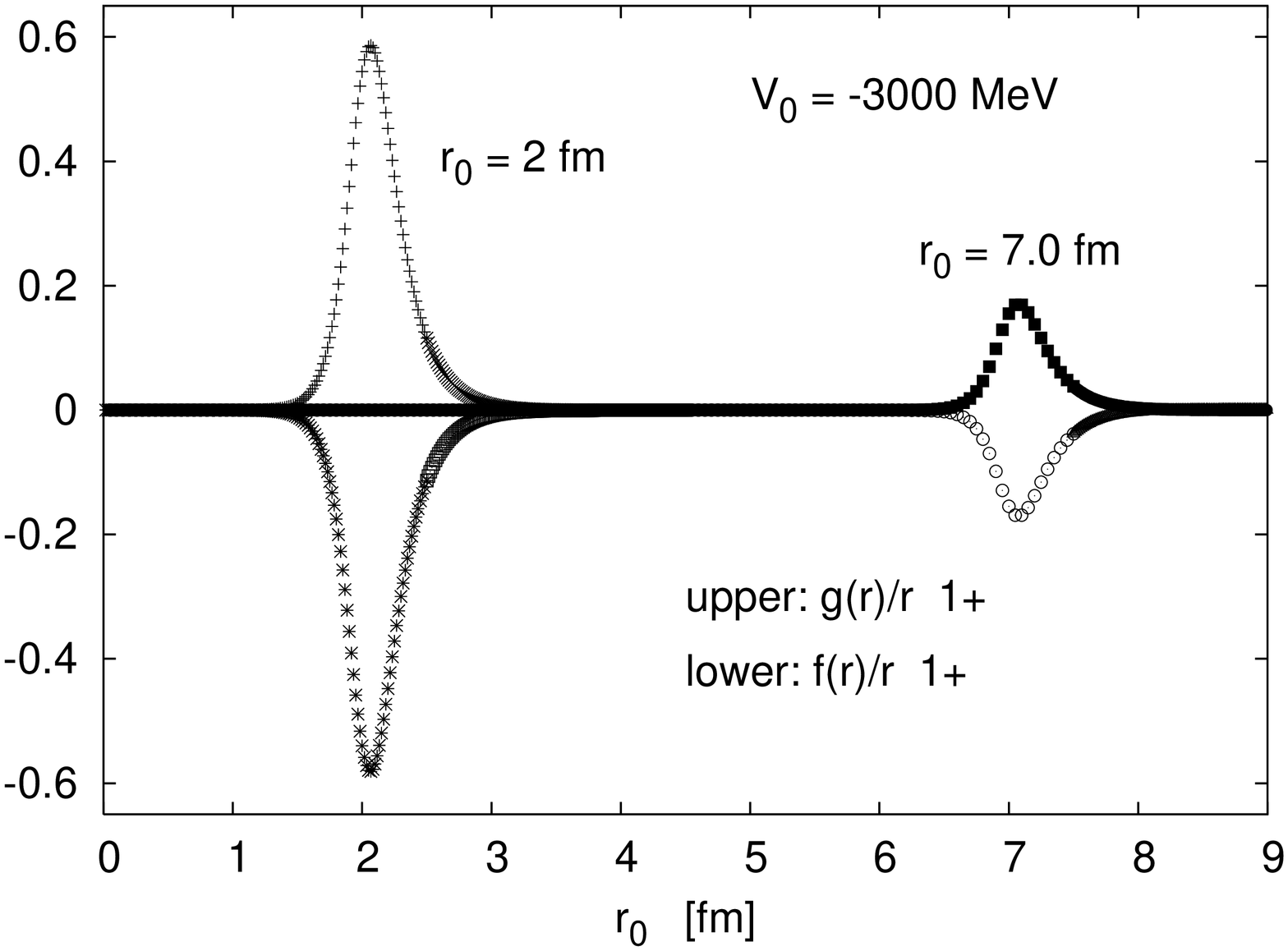}}
\caption{Radial wave functions $g(r)/r$ and $f(r)/r$  for the
$1+$ state of potential wells $V_0$= -3000 MeV and
 $r_0$= 2 and 7 fm. }\label{figure3}
\end{figure}

 Let us now discuss the results from potentials deeper than those
shown in Table 1. In Fig. 10 we focus on the $1+$ state calculated
with different well depths. Plotted are the anti-ncleon fraction
$<f|f>$ and the effective mass
defined as $m^*/m \equiv E_{1+}/(mc^2)$ where $m$ is the free nucleon
mass. Near zero potential, we have $m^*/m \simeq 1$ and $<f|f> \simeq 0$.
As $V_0$ becomes more negative, the former descends and the latter
ascends until $V_0 \simeq -1500$ MeV. Afterwards, its $m*/m$ remains
flat at about 0.05 (corresponding to $E_{1+}\simeq 50$ MeV)
and its $<f|f>$ is flat at about 0.50.
In the figure we show results down to $V_0\simeq-3500$ MeV.
We have tried deeper potentials, in fact  as deep as $V_0=-5400$ MeV,
and  the flatness of these two curves remains unchanged.
It seems that the above  $E_{1+}$ and $<f|f>_{1+}$ are
practically independent of $V_0$ for $V_0 < \sim -1500$ MeV.
 The wave function and energy of this 'flat' state will be referred
to as $\Phi _{1/2}$ and $E_{1/2}$ respectively.

\begin{figure}[h]
\scalebox{0.4}{\includegraphics[angle=-90]{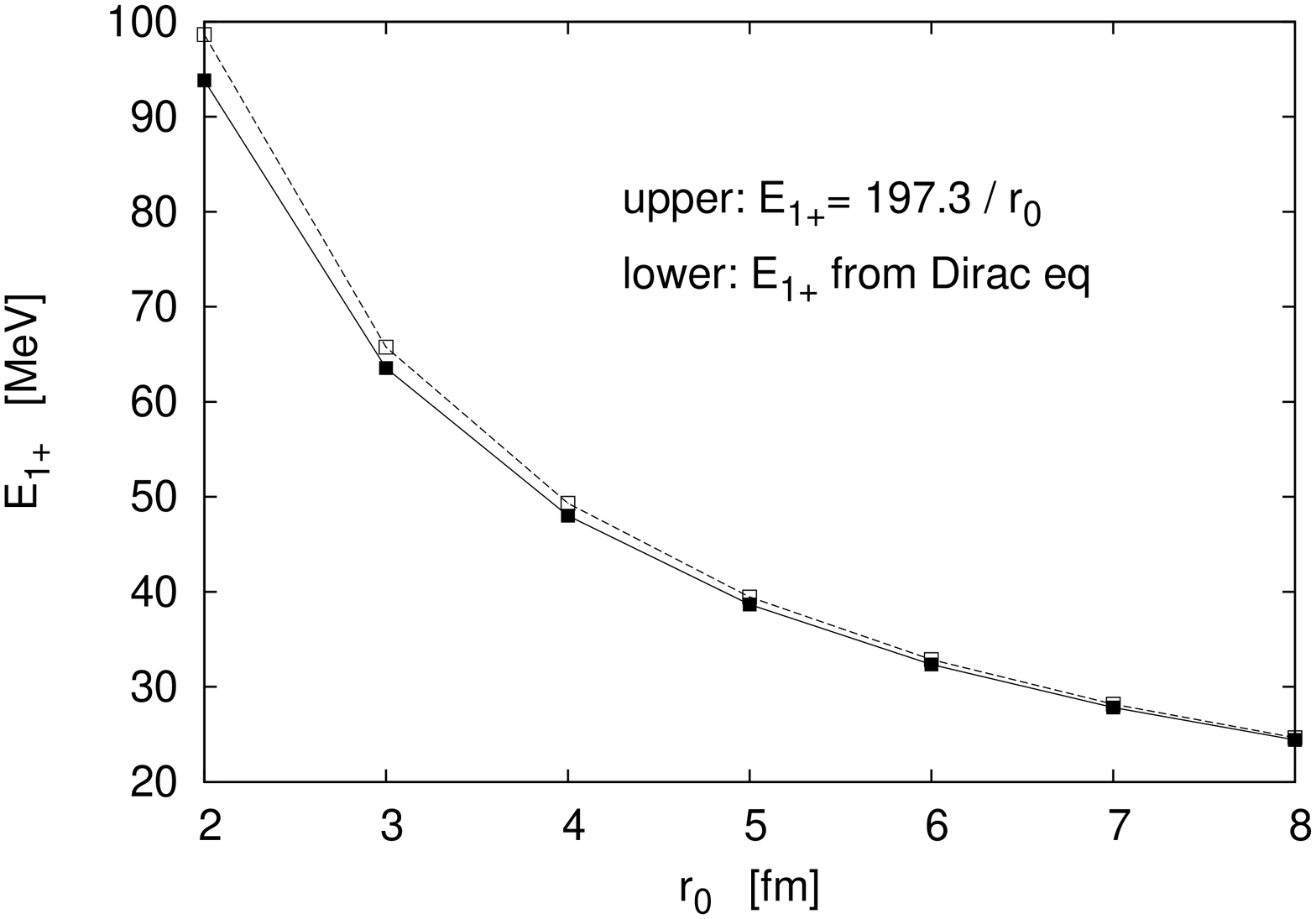}}
\caption{Dependence of the  $1+$ energy on the
range $r_0$ of the WS potential well. Common well depth of
$V_0$=-3000 MeV is used.}\label{figure3}
\end{figure}

  Our results presented above are all obtained numerically.
It would be nice if they can be obtained analytically.
We have tried  but so far have not succeeded.
To partially understand the above 'flatness' behavior
of Figs. 5, 6 and 10,
let us rewrite Eq.(2)  in matrix form as
\begin{equation}
\left[ \begin{array}{cc} -(V_s+mc^2)/\hbar c & \frac{d}{dr}+\frac{k}{r} \\
                         -\frac{d}{dr}+\frac{k}{r}& (V_s+mc^2)/\hbar c
                   \end{array} \right]
\left[ \begin{array}{c} f \\ g
                         \end{array} \right]
=E\left[ \begin{array}{c} f \\ g
                         \end{array} \right].
\end{equation}
Here we do not have the $V_v$ term of Eq.(2)
as we include only the scalar potential $V_s$.

  The diagonal terms $-(V_s+mc^2)$ and $(V_s+mc^2)$ play an important
role in the above equation. When $V_s \simeq 0$,
the wave functions
of the positive- and negative-energy states are dominated
 by their nucleon and anti-nucleon  component $g(r)$ and $f(r)$
respectively. But when $V_s$ is largely negative, say -2000 MeV,
$<f|-(V_s+mc^2)|f>$ becomes postive
while $<g|(V_s+mc^2)|g>$ becomes negative. Then the positive-energy state
is dominated by the anti-nucleon component $f(r)$ while
the negative-energy one by $g(r)$.
  This explains the level crossing
 seen in Fig. 5  for the $1-$ to $2+$ right-to-left
cross over and similarly the $2+$ to $1-$ cross over.

 The middle $1+$ state of Fig. 5 is, however, a special case.
Its energy $E_{1+}$ for $V_0 < \sim -1500$ MeV is not only
nearly independent of $V_0$ but is also
rather small ($\sim 50$ MeV). These properties suggest that
 there must be some cancellation among the
 contributions from $V_s$ to
 this state, leading to a near-zero net contribution.
A possible way to attain such cancellation is to have
either $f=-g$ or $f=g$ where $f$ and $g$ are respectively
the anti-nucleon and nucleon component of the eigenfunction
of this $1+$ state.



 To study this possible cancellation, we rewrite Eq.(10)  as
\begin{equation}
[-(V_s+mc^2)/(\hbar c)+\frac{d}{dr}](f+g) -\frac{k}{r}(f-g)
= E(f-g)
\end{equation}
and
\begin{equation}
[-(V_s+mc^2)/(\hbar c)-\frac{d}{dr}](f-g) +\frac{k}{r}(f+g)
= E(f+g).
\end{equation}
We have used the above equations to calculate the wave
functions (f+g) and (f-g). These calculations can also
be performed equivalently using Eq.(10). We have found
for $V_0 \lesssim -1400$ MeV (1.5$mc^2$ = 1407 MeV),
the wave function (f+g) is very much smaller than (f-g)
for the $1+$ state. We have calculated for this state
the norm
$N_+\equiv <f+g|f+g>$ and
$N_-\equiv <f-g|f-g>$ with the normalization $N_++N_-=1$.
Our results are $(N_+,N_-)$ = (0.00004, 0.99996)
for $V_0=-1400$ MeV and (0.00002, 0.99998) for $V_0$=-1500 MeV.
In both calculations $r_0=4$ fm is used.

Thus for the $1+$ state with $V_0 \lesssim -1500$ MeV,
 we can drop the $(f+g)$ terms relative to the
$(f-g)$ ones in the above equations.  Then Eq. (11)
 gives the simple energy relation
\begin{equation}
E=-k\hbar c  \frac {<g|1/r|g>}{<g|g>}
\end{equation}
for the energy of the $1+$ state. Recall that $k=-1$ (see section 2).
 To calculate the above energy, we need to have $<g|1/r|g>$.

If $<g|1/r|g>/<g|g>$ is independent of $V_0$, then so is the above energy.
In fact we have found that for $V_0 \lesssim -1500$ MeV,
 $(<g|1/r|g>/<g|g>)$ is closely equal to $ 1/r_0 $ and is nearly
independent of $V_0$. Thus the above energy is only weakly dependent
on $V_0$ as indicated by Fig. 10  and Table 1.
 For the above $V_0$ range,
 $g$ is generally
of a Gaussian wave packet shape narrowly peaked around $r_0$.
In Fig. 7 we have displayed such wave functions for $r_0$= 4 fm.
We have found that the wave functions for other
choices of $r_0$ are also of this shape. In Fig. 11 we display two
such wave functions for $r_0$= 2 and 7 fm.
 This special shape of the wave functions renders
 $(<g|1/r|g>/<g|g>) \simeq 1/r_0 $  a good approximation.
With this approximation, Eq. (13) becomes then
\begin{equation}
E \simeq \frac{-k\hbar c}{r_0}.
\end{equation}
The accuracy of this approximation will be discussed later.

We have considered Eq. (11) for $f\simeq -g$. Let us now consider
Eq. (12).  In this situation we can drop the (f+g) terms there
and have
\begin{equation}
\frac{dg}{dr}=-g(V_s+mc^2)/(\hbar c)
\end{equation}
which is a simple differential equation for determining
the wave function $g$.
For the Woods-Saxon potential
of Eq.(8), the above equation gives
\begin{equation}
 g(r) =exp[-\frac{V_0}{\hbar c}[(r-r_0)
 -\delta~ log(1+ e^{(r-r_0)/\delta})] -\frac{mc^2}{\hbar c}r].
\end{equation}
 Note that the above expression does not include the
 normalization constant.
For $V_0 \lesssim -1500$ MeV the  wave function
calculated with the above
equation
 agrees very well with the corresponding result from solving
the full Dirac equation of Eq. (10).

The following approximation is helpful in understanding
the structure of the $g(r)$ wave function.
For $r$ close to $r_0$, the Woods-Saxon potential of Eq.(8)
can be approximated by  a linear potential, namely
\begin{equation}
V_s(r) \rightarrow \frac{V_0}{2}[1 -\frac{r-r_0}{2\mu}].
\end{equation}
Then  Eq.(15) gives
\begin{equation}
g(r)\propto exp[\frac{V_0}{8\mu}(r-r_0)^2]
\end{equation}
which is of  Gaussian form, providing an explanation for the
shape of the wave functions
 displayed in Figs. 7 and 11.

 The parameter $\mu$ of Eq.(17) is determined by fitting
the WS potential
near $r_0$. As an example,  we find $\mu$= 0.144 by fitting
the $r_0$= 4 fm and
$V_0= -2000$ MeV potential
in the range of $r=$ 3.7 to 4.3 fm.
  Then the wave function given by Eq.(18)
is of the form $exp[b(r-r_0)^2]$ with $b=V_0/(\hbar c 8 \mu)=$-8.80.
The same wave function obtained by solving the Dirac equations of Eq. (10)
has been given in Fig. 7. Fitting it by
$\alpha \cdot exp[\beta (r-\gamma r_0)^2]$
gives $\alpha$= 1.07, $\beta$= -8.37 and $\gamma$= 1.004.
Note that the $\beta$ value is
 close to the above $b$ value.

 To estimate the accuracy of the  energy approximation of Eq.(14),
   we have
 compared the energies so obtained with those
from the full calculation of Eq. (10). Some sample
comparisons are given in Fig. 12 where the energies
for the $1+$ state calulated from   Eq.(10) using $V_0=$-3000 MeV and
$r_0$= 2 to 7 fm are compared with those
given by   Eq.(14). It is of interest that the two sets of energies
agree  remarkably well, indicating this  approximation
being highly accurate.
Note that  Eq. (14) is independent of $V_0$. This is consistent
with Fig. 10 where the energies of the $1+$ state are nearly independent
of $V_0$ for $V_0 \lesssim -1500$ MeV.

 For particles of zero rest mass, the momentum-position
uncertainty principle may be written as
~$\Delta \it E ~ \Delta r \simeq \hbar c;~ \it E=pc$.
It is of interest that the energy relation of this
special state as given by Eq. (14) is similar to this
uncertainty principle, suggesting that the nucleon
in this special state has a near-vanishing effective
mass as indicated by Fig. 10.



\section{Summary and discussion}\label{cross}
 In the present work we have solved  the Dirac radial equation
for a nucleon in a scalar Woods-Saxon potential well with
depth $V_0$ and radius $r_0$. For weak potentials such as
 $V_0 \simeq$ -50 MeV, the  positive-energy
states  $\Psi _+$ of this equation have  energies $ \simeq +mc^2$
and  wave functions  dominated by
the nucleon component $g(r)$, as indicated by Figs. 2 and 6.
Similarly the negative-energy states $\Psi _-$ have  energies
close to $-mc^2$ and wave functions dominated by
the anti-nucleon component $f(r)$ as illustrated in Fig. 3.

As $V_0$ turns more negative (deeper), a qualitative reversal of
the wave functions takes place.
Namely, $\Psi _-$ is generally becoming increasingly dominated by
$g(r)$, and so is  $\Psi _+$ by $f(r)$. Examples illustrating
such tendencies are displayed in Figs. 8 and 9.

The lowest positive-energy state $1_+$ is however an exemption,
 being a special intruder state. As indicated by Figs. 5 and 6,
at weak potential strength this state has $E_{1+}\simeq mc^2$
and $<f|\Psi _{1+}>^2 \simeq$ 0. But as $V_0$ becomes
$\lesssim -1500$MeV,
this state is characterized by a special wave function
 $\Psi _{1/2}$
with  constant composition
of $f=-g$ where $g$ and $f$ are respectively its nucleon and anti-nucleon
component and both are narrowly peaked around $r_0$.
In addition, the energy of this state
 approaches a constant energy
$E_{1/2}\simeq \hbar c/r_0$.
These behaviors have been
illustrated in Figs. 5, 6  and 10.
For $V_0 \lesssim -2000$ MeV, our results
suggest that the above state, with wave function $\Psi _{1/2}$
and energy $E_{1/2}$, is the only bound state.
It should be of interest if one could check this special
intruder state  experimentally. We note that
a hypothetical scalar Woods-Saxon potential of very strong
strength has been used in our work
for calculating this and other intruder states.
Is such a strong potential conceivable?



We have estimated that a nucleon near the surface of a
medium-mass black hole (of mass  $\sim 10^3$ solar mass
and radius    $\sim 10^3$ km
\cite{blackholewiki,schutz2009})
would experience
a gravitational potential of approximately
-1400 MeV ($\cong 1.5mc^2$).
There has been evidence for super-massive black holes
such as the $4.3\times 10^6$ solar-mass black hole
in the Milky Way
\cite{blackholewiki,schutz2009}.
They may produce similar or stronger gravitational potential.
Thus, the special nucleons indicated by this work
may be present near such black holes.
\footnote{We have also made
a similar estimate for a neutron star of mass 2.1$M_{\odot}$
\cite{anton13} and radius $\sim 11$km \cite{paenghls16}.
The resulting potential is about -270 MeV. Thus we do not
expect the presence of such special nucleons
 near this neutron star,
as indicated by Fig.10.}
This  needs of course further study. With the presence of
such strong gravitational field we may need to use
a Dirac equation in curved spacetime
\cite{mukhopad2000,carroll2004}.
whereas we have been
using the Dirac equation in flat spacetime.
It would indeed be of interest to obtain
an estimate of the effects introduced by curved
spacetime.



Let us end by quoting a paragraph written by Gerry \cite{brown2010}:

``One of the authors, Gerry Brown, arrived at Princeton in early
September, 1964. The next morning,
as he came to the Palmer Physics Laboratory,
Eugene Wigner, who just preceded
him, opened the door for him (It was a real contest to get ahead of
Eugene and open the door for him which very few succeeded in doing.)
Eugene asked Gerry, as we went into the building, what he planned to
work on. `I plan to work out the nucleon-nucleon interaction in
nuclei.' Eugene said that it would take someone cleverer than him,
to which Gerry replied that they probably disagreed what it meant to
`work out'. Gerry wanted to achieve a working knowledge, sufficiently
good to be able to work out problems in nuclear physics.''

    In fact, it is our  hope that the present calculations
 may serve as a warm-up exercise for 'working out'
  the Dirac equation for
nucleons experiencing extremely strong scalar potential fields
and its possible connection with dense stellar objects.

{\bf Acknowledgement:} We thank Edward Shuryak, Ismail Zahed
and Jeremy Holt for many helpful discussions.
This work is  supported by
the US Department of Energy under contract DE-FG02-88ER4038.
S. Shu is supported by the China Scholarship Council.

\begin{figure}[h]
\scalebox{0.43}{\includegraphics[angle=-90]{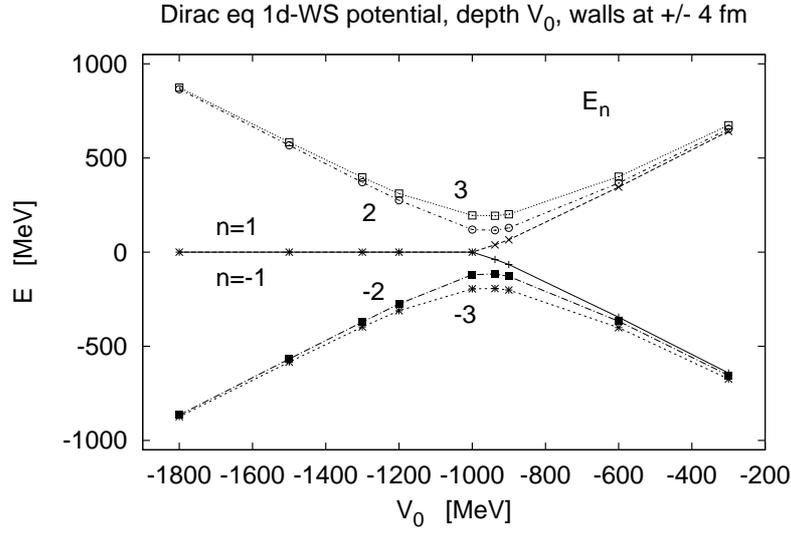}}
\caption{Energies  of  Dirac equation for a nucleon
in a one-dimensional scalar WS potential. A similar  plot
 for the spherical WS potential used earlier is shown
in Fig. 3.}
\end{figure}

\begin{figure}[h]
\scalebox{0.43}{\includegraphics[angle=-90]{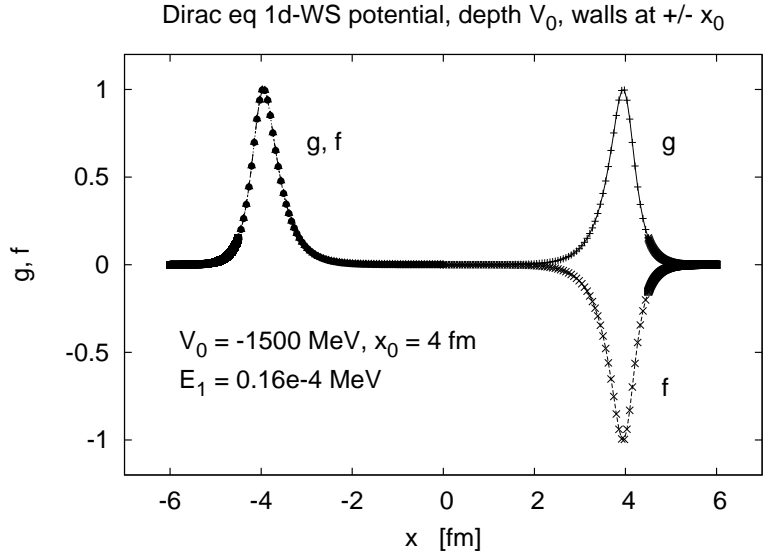}}
\caption{ Wave function of a one-dimensional half-nucleon intruder state.
Similar wave functions for the spherical WS potential
used earlier are shown in Figs. 7 and 11.}
\end{figure}

\begin{figure}[h]
\scalebox{0.43}{\includegraphics[angle=-90]{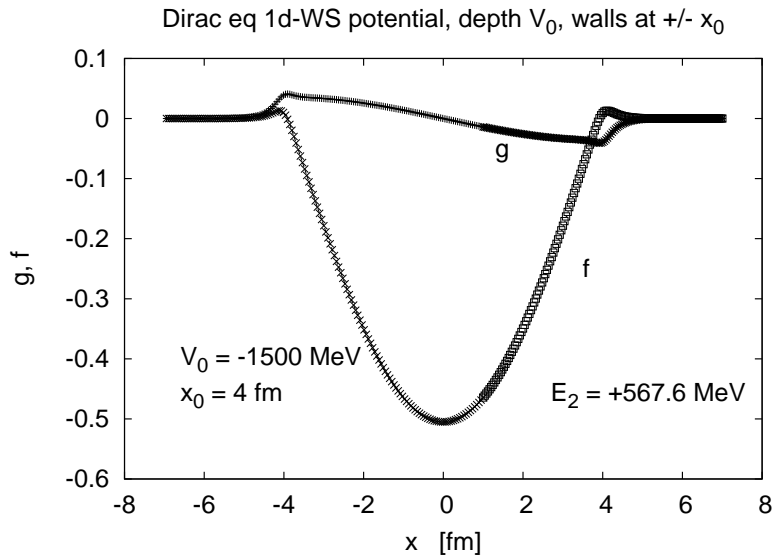}}
\caption{Wave function of a one-dimensional positive-energy state
with $f$ dominance. Similar wave functions
for the spherical WS potential used earlier are shown in Fig. 9}
\end{figure}

 {\bf Appendix:}
 In this Appendix
 the nucleon-anti-nucleon intruder
states mentioned earlier  are studied using a  simpler
and  more transparent  model.
As indicated by Eq.(5), our earlier calculations are based on
a model where  a nucleon is placed
in a spherical Woods-Saxon (WS) potential.
 When the depth of this potential
is deeper than a certain value,  the ground state of the nucleon
becomes a special
half-nucleon intruder state.
One might ask if this holds true also for a one-dimensional
Woods-Saxon potential.

 To answer this question, we have repeated our earlier calculations
using  a one-dimensional WS potential
\begin{equation}
V_s(x)= V_0[1+ exp(\frac{~xS(x)-x_0~}{\delta})]^{-1};~S(x)=x/|x|.
\end{equation}
 The corresponding Dirac equation is
\begin{equation}
\left[ \begin{array}{cc} A & B \\
                         -B& -A
                   \end{array} \right]
\left[ \begin{array}{c} f \\ g
                         \end{array} \right]
=E\left[ \begin{array}{c} f \\ g
                         \end{array} \right]
\end{equation}
with
\begin{equation}
A=-(V_s(x) + mc^2)/(\hbar c), ~~ B=\frac{d}{dx}.
\end{equation}
The solutions of this equation are paired, namely
\begin{equation}
E_{n+}=-E_{n-};~~(f_{n+},g_{n+})=\pm (g_{n-},f_{n-})
\end{equation}
for all n. Eq.(20) can be rewritten as
\begin{eqnarray}
(A-B)(A+B)(f+g)& =& E^2 (f+g) \nonumber \\
(A+B)(A-B)(f-g)& =& E^2 (f-g).
\end{eqnarray}

  The energies of these equations have been calculated as
 illustrated in Fig. 13.
(A common value of $\delta$ = 0.1 fm is used in all calculations.)
 We note that the energies are paired, in agreement with
Eq. (22).

 As also shown in Fig. 13, there is a clear level crossing behavior near
$V_0 \simeq -1000~ MeV$.
For potentials shallower than this
value,
the wave functions for the $E_{n+}$  and $E_{n-}$ states are dominated
respectively by $g$ and $f$.  For potentials deeper,
the energies of the $E_{1+}$ and $E_{1-}$ states are special, both
being practically  zero and independent of the potential depth,
and their wave functions  all have a special
structure of either $f=-g$ or $f=g$ as illutrated in Fig. 14.
Also for this potential range, the wave functions for
the ($E_{n+},n>1$) states are dominated by $f$ as illustrated
in Fig. 15. Similarly those
for the ($E_{n-},n>1$) states are dominated by $g$.

  According to Eq. (20), the wave function $(f, g)$
as a whole has a
'mixed parity' structure, in the sense that each of $f$ and $g$
is a function of definite parity but their parities are
 opposite.
 This is seen in Fig. 14 where
$g$ has positive parity while $f$ has negative parity.
Similar parity structure is also seen in Fig. 15.
We note that Eq. (23) provides a convenient way for obtaining
approximate  wave functions near the
 walls and and near the center of the potential.
This has been helpful in explaining the shapes of the
wave functions of Eq.(20).

 In short, the main features of the nucleon-anti-nucleon
intruder states obtained from the spherical WS potential
are well reproduced by the results from
 the one-dimensional WS potential.


\end{document}